IAPS | Institute for AI Policy and Strategy

September 2024

# Mapping Technical Safety Research at AI Companies

A literature review and incentives analysis

**AUTHORS**

Oscar Delaney – Research Assistant

Oliver Guest – Research Analyst

Zoe Williams – Acting Co-Director

# Executive Summary

As artificial intelligence (AI) systems become more advanced, concerns about large-scale risks from misuse or accidents have grown. This report analyzes the technical research into safe AI development being conducted by three leading AI companies: Anthropic, Google DeepMind, and OpenAI.

We define "safe AI development" as developing AI systems that are unlikely to pose large-scale misuse or accident risks. This encompasses a range of technical approaches aimed at ensuring AI systems behave as intended and do not cause unintended harm, even as they are made more capable and autonomous.

We analyzed all papers published by the three companies from January 2022 to July 2024 that were relevant to safe AI development, and categorized the 80 included papers into nine safety approaches.[1] Additionally, we noted two categories representing nascent approaches explored by academia and civil society, but not currently represented in any research papers by these leading AI companies. Our analysis reveals where corporate attention is concentrated and where potential gaps lie (Table 1).

Some AI research may stay unpublished for good reasons, such as to not inform adversaries about the details of safety and security techniques they would need to overcome to misuse AI systems. Therefore, we also considered the incentives that AI companies have to research each approach, regardless of how much work they have published on the topic. In particular, we considered reputational effects, regulatory burdens, and to what extent the approaches could be used to make the company's AI systems more useful.

We identified three categories where there are currently no or few papers and where we do not expect AI companies to become much more incentivized to pursue this research in the future. These are model organisms of misalignment, multi-agent safety, and safety by design. Our findings provide an indication that these approaches may be slow to progress without funding or efforts from government, civil society, philanthropists, or academia.

---

[1] Because of issues with our code, an earlier version of our research incorrectly excluded some relevant papers. We apologize for the error. Our corrected paper was finalized on September 25th 2024 and the original version can still be accessed for archival purposes on arXiv. The issues are described in more detail on our GitHub repository.

**Table 1: Amount of public research in different categories from leading AI companies**

| Area | Proportion of relevant papers[2] |
|---|---|
| **Enhancing human feedback**<br>Developing better ways of incorporating human preferences when training advanced AI models in cases where people might struggle to give adequate feedback on AI outputs. | 39% |
| **Mechanistic interpretability**<br>Developing tools to convert model weights into useful higher-level human concepts describing the model's beliefs and reasoning processes. | 24% |
| **Robustness**<br>Improving the worst-case performance of AI systems even on anomalous inputs, reducing the likelihood of unpredictable and unintended behaviors in novel situations. | 13% |
| **Safety evaluations**<br>Assessing whether an AI system possesses dangerous capabilities, to inform decisions about mitigations and whether the system is safe enough to deploy or continue to train. | 11% |
| **Power-seeking tendencies**<br>Understanding whether and how AI systems display power-seeking tendencies, and investigating methods to inhibit such tendencies. | 4% |
| **Honest AI**<br>Ensuring that AI systems accurately communicate their beliefs and reasoning, making it easier to detect harmful goals and plans in highly capable systems. | 4% |
| **Safety by design**<br>Pioneering novel approaches to building intrinsically safe AI systems, such as with formal proofs that systems will behave in certain ways. | 3% |
| **Unlearning**<br>Deliberately making models less capable for some dangerous tasks. | 3% |
| **Model organisms of misalignment**<br>Creating simple demonstrations of AI deception or other concerning behaviors, and testing whether proposed safety techniques work on these examples. | 1% |
| **Multi-agent safety**<br>Understanding and mitigating risks from interactions between AI systems.[3] | 0% |
| **Controlling untrusted AIs**<br>Techniques to make AI models less dangerous, even if those models are "misaligned," i.e., inclined to act in a way that the developer does not intend. | 0% |

---

[2] "Relevant papers" refers to research published by Anthropic, (Google) DeepMind, and OpenAI between January 2022 and July 2024 (inclusive) that met our inclusion criteria.

[3] Although we could not find papers in scope that are specifically about safe multi-agent safety, there are papers about multi-agent interactions in general, particularly from DeepMind (for example Agapiou et al., 2023).

# Table of Contents

# Introduction

There are significant concerns from governments, civil society, and AI companies[4] that advanced AI systems may pose large-scale risks from misuse or accidents (The White House, 2023; Bengio et al., 2024; Center for AI Safety, 2023).

In this paper, we explore a particular subset of the work being done to reduce large-scale AI risks. Specifically, we investigate AI companies' technical research into safe AI development, and what research approaches are more common.

We use "safe AI development" to mean developing AI systems that are unlikely to pose large-scale misuse or accident risks. This is related to, but broader than, the concept of AI 'alignment,' which generally refers to making AI systems act as intended by the developer (Guest, Aird, and Ó hÉigeartaigh, 2023, pp. 33–34). Safe AI development is just one topic where technical research might be needed to reduce AI risks. Other types of potentially valuable research include:

- **Reducing risks posed by AI systems that have already been deployed** such as by making "deployment corrections" (O'Brien, Ee, and Williams, 2023).
- **Technical AI governance**, i.e., using technical analysis and tools to support AI governance. For example, using technical measures to make it harder to steal high-risk AI models (Reuel et al., 2024).
- **Systemic AI safety**, i.e., reducing AI risks by focusing on the contexts in which AI systems operate (UK AI Safety Institute, 2024b). In an influential paper, Hendrycks et al. (2022) focus particularly on using machine learning techniques to increase systemic AI safety. For example, using AI systems to increase cybersecurity to defend against AI-enabled cyberattacks.

We also comment on the incentives that AI companies have to do different types of research. We use this incentives analysis to make some tentative predictions about future changes in the sorts of safety research companies will do.

Understanding what research AI companies are doing, and will do in future, is valuable for allocating research funding from other actors. If funders do not take into account what research AI companies are likely to do, then they might waste resources by duplicating efforts. Additionally, they might ideally fund research that makes different 'bets' to AI companies' research, making a diversified portfolio of safety investments for greater defense in depth (Barnes, 2024, pp. 100–101). That said, our paper cannot be used on its own to identify promising funding opportunities; we leave other relevant considerations (noted below) out of scope. We plan to make recommendations for government and philanthropic funding that consider these factors in follow-up work.

---

[4] By "AI companies", we mean the companies that are developing AI systems. We particularly focus on companies that are training very compute-intensive models, i.e., models that require computation worth tens of millions of dollars to train. Examples of such companies include Anthropic, Google DeepMind, and OpenAI. Many experts have linked the most severe AI risks specifically to compute-intensive models (Bengio et al., 2024).

> ### Additional considerations for research into safe AI development
> This paper explores AI companies' likely research focus. When deciding what research would ideally be supported by other actors, such as governments or philanthropists, several additional considerations are beyond our scope:
> - **How promising is the research approach?** Everything else being equal, funders and researchers should focus on approaches that are more likely to be fruitful.[5]
> - **Might the research bring additional benefits?** Research could be helpful both for reducing large-scale AI risks and for achieving other desirable goals.
> - **Might the research cause harm?** Researchers and funders should be more reluctant to support research that does harm as well as bringing benefits.[6]
> - **Is the relevant institution in a good position to do or fund this research?** Specific institutions will have expertise or structures that make them particularly well-suited to support particular kinds of research.[7]

In the following section we describe our method. We then present our results, going through the different approaches to safe AI development, sorted by number of papers. We conclude with thoughts on what categories might see more research from AI companies in future.

---

[5] Predicting research success in advance is difficult. As a result, funders should often be willing to invest in 'moonshots' that seem unlikely to work, but would have an outsized impact if they do. Additionally, investing in addressing risks not addressed by other approaches is especially valuable.

[6] See Guest, Aird, and Ó hÉigeartaigh (2023) for an overview of ways in which efforts to support alignment research might counterintuitively increase large-scale AI risks.

[7] For example, academia might be particularly well-suited for research into mechanistic interpretability. Improving interpretability techniques does not require using the largest models (which are sometimes out of reach of academic researchers), and is amenable to publishing papers (Kästner & Crook, 2023; Zimmermann, Klein, and Brendel, 2024), consistent with academic incentives.

# Methods

## Defining approaches to safe AI development

Our first step was identifying distinct categories of research within safe AI development. As described above, this means focusing on technical research that could be used to develop AI systems that are unlikely to pose large-scale accident or misuse risks. We carried out a literature review of research agendas and taxonomies that are relevant to safe AI development[8] and consulted multiple experts in the field. We then collated research approaches into 11 key categories.

## Quantitative analysis of published research

Our primary method was to determine how many papers by leading AI companies about safe AI development are in each of the categories that we identified. The papers that we collected, and our categorizations of them, are available on a [Google Sheet](). Code and details about our method is on [GitHub]().

We had three inclusion criteria for papers:
- **Strongly associated with Anthropic, (Google) DeepMind, or OpenAI.**[9] We operationalized this as research that is published on those companies' websites or research where the first author lists one of these companies as an affiliation. We focused on these three companies because they are arguably the leading developers of advanced AI.[10] As a result, they are particularly well-placed to do research into safe AI development, as well as organizations to whom this research is particularly relevant.
- **Published between January 2022 and July 2024.** We chose this time period to balance collecting a large number of papers with being up-to-date.
- **Relevant to safe AI development.** To make this judgment we reviewed the paper's title and abstract, and, if we were uncertain, the rest of the paper.

---

[8] The agendas and taxonomies were framed in slightly varying ways, e.g., sometimes in terms of "alignment" or "ML safety". The main sources that we consulted were Hendrycks et al., 2022; Open Philanthropy, 2022; Critch & Krueger, 2020; Ji et al., 2024; Anwar et al., 2024; Toner & Acharya, 2022; and Amodei et al., 2016.

[9] DeepMind has been a part of Google since 2014. In April 2023, it was merged with Google Brain, another AI group within Google, and renamed Google DeepMind (Murgia, 2023). We use "(Google) DeepMind" to refer to DeepMind (but not Google Brain) up to April 2023 and Google DeepMind from April 2023.

[10] For instance, they have carried out the largest publicly known training runs and are among the companies responsible for the most citations in AI research (Cottier, 2023). Additional points of evidence include that these companies currently seem to have the best chatbot products (Chiang et al., 2024) and that they are referred to as the primary artificial general intelligence companies in other research, such as Schuett et al. (2023, p. 3). Other companies that would be relevant by some of these criteria include Meta, Microsoft, and non-DeepMind parts of Google.

We looked for papers in two places:
- **Company websites:** We manually reviewed posts on the websites of the three companies to see whether they linked to research that meets our inclusion criteria.
- **ArXiv:** ArXiv is a preprint server where machine learning research is often published. We programmatically collected papers on arXiv that were from the relevant time frame, had specific terms in the title relating to safe AI development, and where the first author listed an affiliation at one of the three companies. We manually reviewed the resulting papers for inclusion in our dataset.[11]

Our final dataset had 80 papers. We manually reviewed the title and abstract of each of these papers to determine the right category.[12] Some papers are relevant to several of the research categories, in which case we picked the single best fit.

A limitation of our method is that it does not include unpublished research, or research that AI companies share publicly but not as a paper. This is potentially a significant limitation, as there are several reasons why AI companies might not publish their research into safe AI development.[13] For example, research into safe AI development can sometimes be used to make AI systems both safer and more capable (Guest, Aird, and Ó hÉigeartaigh, 2023, pp. 15–20; Brady, 2024; Khlaaf, 2023). Making results from such research accessible to competitors might put companies at a commercial disadvantage (Albergotti, 2024). Making results accessible to foreign governments might have national security implications. Additionally, companies might have mundane reasons not to publish the research, such as not wanting to go through the process of writing a paper.

## Qualitative analysis of companies' incentives

We also did a qualitative analysis of the corporate incentives that AI companies have to do research in the various categories. This helps to adjust for the fact that some research will not be published. It might also be helpful for making predictions about how research by AI companies will change over time. We focus on three kinds of incentives that AI companies have to do research into safe AI development.[14]

---

[11] Because of issues with our code, an earlier version of our research incorrectly excluded some relevant papers from arXiv. We apologize for the error. Our corrected paper was finalized on September 25th 2024 and the original version can still be accessed for archival purposes on arXiv. The issues are described in more detail on our GitHub repository.

[12] As explained in more detail on GitHub, we also used a language model to check our categorizations. In some cases, we updated our categorization based on this step.

[13] This may vary by research category as well. For instance mechanistic interpretability research is relatively unproblematic to publish, whereas some types of adversarial robustness research are safer to not publish, lest adversaries learn how to defeat safety strategies employed.

[14] It may also be that the values of the employees or leadership of companies causes them to do more safety research than corporate incentives would suggest. Indeed, the three leading AI companies were each founded with reducing large-scale AI risks as important motivations (Perrigo, 2023a; Matthews, 2023; Metz, 2016). However, it is

## Incentives for AI companies to do safe AI development research

**Reputation:** Companies with better reputations for safe AI development will probably find it easier to attract the best talent and the most customers. Recent surveys have shown majorities of respondents are concerned about risks from AI and support safety efforts (Department for Science, Innovation & Technology, 2023b; Pauketat, Bullock, and Anthis, 2023). Safe AI development might also reduce the downside risks of a company's reputation being harmed if its model is involved in a large-scale incident, either through misuse or accident. Different research categories in safe AI development might contribute differently to the reputations of AI companies. For example, some categories might be of more interest to the public.

**Regulation:** AI companies might do research into safe AI development so that they can comply with regulations. These might include regulations that directly regulate safety and security risks, or regulations about other topics that touch on safety and security risks. For example, many jurisdictions are considering rules that would require AI decision-making to be "explainable" in important contexts (Information Commissioner's Office and The Alan Turing Institute, 2022, pp. 10–15). The technical developments required to implement this might also be helpful for preventing AI accidents, such as by helping us better understand how AI systems make decisions.

**Usefulness:** Some research into safe AI development might have a co-benefit for AI companies via also improving the usefulness of the company's AI systems.[15] For example, reinforcement learning from human feedback was initially developed as a safety technique, but also makes AI systems more likely to respond helpfully to user queries (Casper et al., 2023).

---

unclear how values around reducing large-scale AI risks would translate into decisions about which research categories to prioritize. Moreover, companies may become increasingly profit-driven as the financial stakes of AI development become larger. Therefore, we did not consider companies' values in our analysis.

[15] Indeed, there is no clear boundary between safety and capabilities research; some work could fall in a gray area of being both.

# Results and Analysis

Figure 1 summarizes our results for the 80 papers that met our inclusion criteria. The remainder of this section presents our detailed findings for each research category.

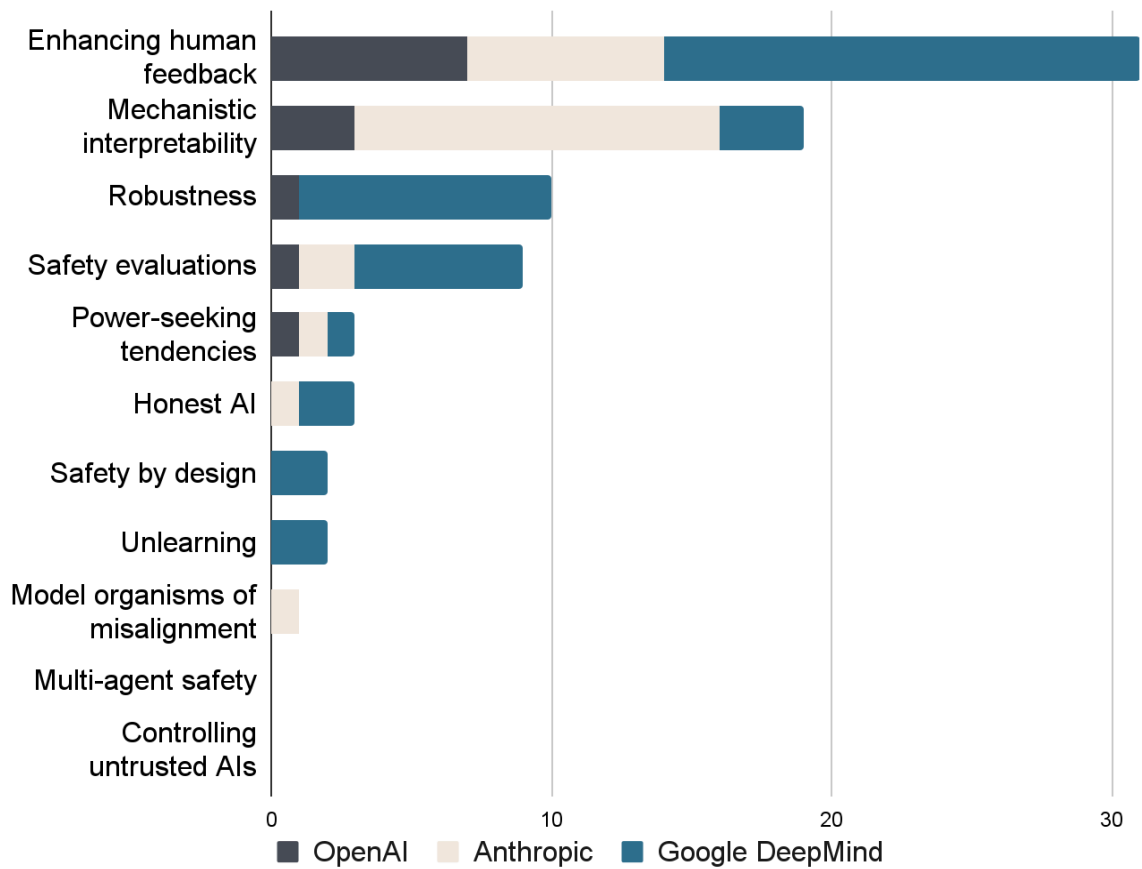

*Figure 1: Distribution of safe AI development research papers by category and by select companies, from January 2022 to July 2024.*

## Enhancing human feedback (31 papers)

Research on enhancing human feedback aims to improve the use of human preference data in training AI systems to make decisions humans would approve of, if they knew all the relevant details. A key technique in this category is Reinforcement Learning from Human Feedback (RLHF), which uses human preferences to fine-tune models (Christiano et al., 2017). RLHF updates model weights to increase the likelihood of producing outputs that human raters would judge as helpful.

While RLHF has proven effective for current systems, it faces a significant challenge known as the "scalable oversight" problem (Bowman et al., 2022; Casper et al., 2023). As AI capabilities grow, it becomes increasingly difficult for humans to effectively evaluate AI outputs they may not fully understand. Additionally, gathering sufficient human feedback might become prohibitively expensive and time-consuming at the scale required for training advanced AI systems.

To address these challenges, researchers are exploring several more advanced techniques:

1. **Reinforcement Learning from AI Feedback** (RLAIF): This approach uses AI systems to evaluate outputs against predefined rules or principles, reducing direct reliance on human judgment. For example, researchers have run trial experiments using one language model to evaluate another (Perez et al., 2022), including where the evaluator language model is far less capable than the evaluatee model, to simulate a future scenario where humans and less capable AIs are evaluating superhuman AI models (Burns et al., 2023). Constitutional AI is another example of this method, where AI models themselves evaluate outputs against the standards of a fixed human-given 'constitution' outlining desired behavior (Bai et al., 2022).
2. **Debate**: This approach gets two AI instances to make arguments and critique each other's reasoning, with a human judge making the final decision (Irving, Christiano, and Amodei, 2018). By having AIs surface key considerations, this approach aims to make human evaluation more efficient and effective, even for complex topics.
3. **Iterated Distillation and Amplification** (IDA): A human assisted by many copies of a trusted, moderately capable AI system will usually perform better than an unaided human at some task. A new, more powerful AI system could then be trained to imitate the behavior of the earlier human + AIs team. Iterating this process could result in AI systems that are both very capable and aligned (Cotra, 2018).

**Corporate incentives**

- Reputation: RLHF and RLAIF are reasonably effective at ensuring that chatbots do not respond to users offensively or insensitively (Ouyang et al., 2022; Bai et al., 2022), making reputation-harming incidents less likely. Debate and IDA are too early in their development to be helpful for currently deployed AI systems, meaning that they currently have less reputational benefit.
- Regulation: Similar to above, RLHF and RLAIF may be helpful for making chatbots less likely to produce outputs that may be prohibited by regulations, such as hate speech or advice on how to commit crimes (Bai et al., 2022).
- Usefulness: The largest benefit from techniques such as RLHF is that AI systems fine-tuned with these methods are much more helpful to interact with (Christiano et al., 2017; Ouyang et al., 2022). This is because rather than, for instance, naively predicting the next token, they will interpret the human's input in a more 'sensible' way and respond to the explicit or implied request. Some kinds of work to enhance human feedback, such as work that builds on RLHF,

> can also yield beneficial outcomes quickly. Techniques that still require years of research would be less attractive to corporate decision-makers.

Although we expect enhancing human feedback overall to benefit from significant research efforts by AI companies, there is a lot of variation within this category. Enhancing human feedback includes techniques that range from being readily implementable and reasonably well-understood (such as RLHF) to techniques that are much more speculative and mostly not yet ready to be used (such as debate). We expect that companies will focus on the former part of this spectrum because pay-offs will be quicker and more likely.

## Mechanistic interpretability (19 papers)

Advanced AI models are like black boxes to us; even though we can examine the inner workings and numerical values that define the model, this doesn't provide human-understandable insights into how the AI reasons or what it believes. Mechanistic interpretability attempts to render AIs intelligible by developing tools for converting model weights into useful higher-level human concepts describing model capabilities, behaviors, and beliefs (Bereska & Gavves, 2024; Räuker et al., 2023). Powerful interpretability tools would allow us to know better when models have dangerous capabilities, what the goals of models are, and whether they are being deceptive (Olah, 2023). This would better inform decisions about whether it is safe to deploy a given model, or what safety features need to be added. Indeed, regulations could be implemented mandating that AI systems cannot be deployed without guarantees, such as from interpretability, that they do not behave in deceptive ways (Clymer et al., 2024, pp. 32-35). Interpretability could also eventually be useful to test the effectiveness of various technical alignment approaches, by examining the internals of models trained with safety techniques and noting whether any concerning features, such as power-seeking propensities, are present (Ji et al., 2024).

Templeton et al. (2024) provide an illustrative example of how mechanistic interpretability research might contribute to reducing AI risks. The authors extracted high-level "features" from a medium-sized language model, identifying where various concepts are stored in the model. These include concepts that are relevant to safety, such as 'unsafe code', 'bias', 'deception', and 'power-seeking'. Templeton et al. demonstrate that model behavior can be changed by artificially up- or down-regulating certain features. However, they acknowledge that their current results only demonstrate the plausible usefulness of this approach, and that significant limitations and open questions remain before interpretability tools could definitively improve safety.

> **Corporate incentives**
>
> - **Reputation:** Progress on mechanistic interpretability will likely be less visible to most customers, at least in the near future, so reputational effects will probably be weak here. Conversely, impressive interpretability work will be legible to technical AI talent, helping with hiring. Alternatively, it may be that the focus on interpretability is more peculiar to the values and history of Anthropic as a company—they produced 12 out of the 17 papers in this category—rather than general corporate incentives.
>
> - **Regulation:** Multiple jurisdictions are considering rules that would require AI decision-making in certain contexts to be explainable, which would likely require significant advances in interpretability (Nannini, Balayn, and Smith, 2023). Governments may also want to impose specific rules on internal features of AI decision-making systems, for instance, that driverless cars should not consider demographic details of affected pedestrians when modeling crash contingencies. Without advances in interpretability, it might be difficult for companies to demonstrate that their systems comply with this regulation. Thus, AI companies might be worried that work on interpretability would accelerate the introduction of burdensome regulation that depends upon interpretability. Alternatively, if such regulations are forthcoming anyway, an AI company that develops a lead in understanding its models may be better placed to compete in a tighter regulatory environment, incentivizing increased investment into interpretability.
>
> - **Usefulness:** Better mechanistic interpretability would give companies deeper insights into their models, and thereby accelerate the science of deep learning, making it easier to design more useful AI systems (see eg Poli et al., 2023). The economic benefits of better interpretability are indirect—we do not know which capabilities will be unlocked or research directions pioneered—but could be significant, as it is harder to make improvements to poorly understood systems.

## Robustness (10 papers)

AI systems often perform considerably worse on inputs that differ greatly from their training data. For example, facial recognition systems are less accurate for groups that are underrepresented in the training data, such as people of color (Buolamwini & Gebru, 2018). Work on robustness aims to ensure that AI systems maintain minimum performance standards even on inputs that are unprecedented for the system. Increasing robustness could reduce large-scale AI risks in several ways (Hendrycks et al., 2022; Gleave, 2023), including:

- Reducing the likelihood of AI systems behaving in unpredictable and unintended ways when they are in new situations.
- Making AI systems less vulnerable to adversarial attacks from humans, i.e., attempts to get AI systems to act in ways not intended by their developers by cleverly tailoring inputs (Zou et al., 2023b). This could reduce misuse risks, such as by making it harder for users to elicit information about designing bioweapons.
- Reducing vulnerability to adversarial attacks from other AI systems.[16]

One proposal to increase robustness is to use adversarial training—generating training data that deliberately seeks to cause model failures, and updating the model to adapt to these inputs—to ensure that models can withstand adversarial attacks post-deployment (Ziegler et al., 2022). Other work seeks to ensure that the patterns a model learns to rely on in a training dataset do not mis-generalize when exposed to a novel environment where those patterns no longer hold (Armstrong et al., 2023). For instance, if a driverless car is trained on data where a stop sign is always of a certain color or shape, but then it encounters a novel stop sign on the road, its learned proxy of 'stop when you encounter a red octagon' may fail, causing an accident.

As an additional approach, developers could aim to ensure that systems that use AI have a "fail-safe" process that they use when the AI is in an unusual situation and so at higher risk of failing. For example, chatbot products could replace the model's text output with a simple message declining to comment, or self-driving cars could bring the car to a controlled stop. That said, identifying cases where a model is an unusual situation remains an unsolved problem (Hendrycks et al., 2022, 5; Rudner & Toner 2024).

**Corporate incentives**
- Reputation: Improved robustness will make AI systems more attractive to deploy as companies do not want their systems to respond unpredictably to anomalous situations. For instance, Microsoft's reputation was harmed during the initial launch of Bing Chat, where an alter-ego called 'Sydney' emerged and threatened users (Perrigo, 2023b). This would be avoidable by a more robust AI system.
- Regulation: Some domains require very high reliability for an AI system to be useful, such as surgical robotic systems, driverless cars, and military AI

---

[16] This is important because many proposals for ensuring that AI systems behave as intended involve AI systems supervising and checking the work of other AI systems. This approach would be unreliable if the supervised AI systems can manipulate the supervising AI systems.

> systems.[17] In these cases, regulation may place a stringent burden of proof on AI companies to demonstrate the robustness of their systems across all plausible scenarios.
> - **Usefulness:** The primary way in which robustness might make AI systems more useful is by allowing them to be used in higher-stakes contexts. In some cases, robustness improvements will reduce peak performance, as the system becomes more conservative to handle a wider range of inputs. But this will often be a valuable trade-off; robustness work will be key to further commercializing AI systems.

Robustness research may be especially likely to be unpublished. Robustness research is often aimed at making models more robust to attacks by adversaries; sharing such research might help those adversaries. As a result, robustness may be even less neglected than is indicated by the number of papers.

## Safety evaluations (9 papers)

Safety evaluations are empirical assessments of whether an AI system has dangerous properties. Shevlane et al. (2023) describe two kinds of safety evaluations for large-scale AI risks.[18] Dangerous capabilities evaluations assess whether an AI system has offensive capabilities, such as the ability to design biological or cyber weapons (Li et al., 2024) and/or capabilities that would allow a misaligned AI system to better evade human oversight, such as the ability to copy its model weights onto a new server (Kinniment et al., 2024).[19] Alignment evaluations assess the propensity of AI systems to apply their capabilities for harm, such as because they are pursuing goals that their users did not intend.

Safety evaluations are commonly used and proposed as a way to reduce large-scale AI risks. A key example is "responsible capability scaling" (Department for Science, Innovation & Technology, 2023a). Anthropic, OpenAI, and Google DeepMind have all published policies and protocols for how they will monitor dangers from their AI systems, and respond with appropriate mitigations as they scale to more capable models (Anthropic, 2023; Dragan, King, and Dafoe, 2024; OpenAI, 2023). Government regulation increasingly also involves model safety evaluations. For example, the EU AI Act imposes a requirement for "general-purpose AI systems" to undergo model safety evaluations—though the detail of what this means still needs to be defined by standard-setting organizations (European Parliament, 2024; Heikkilä, 2024; Pouget, 2023). The Biden AI Executive Order also requires AI companies to share with the US government the results from safety evaluations of the most advanced AI systems (The White House, 2023). Safety evaluations are also a key focus of government research into safe AI development. In particular, the UK AI Safety Institute is developing safety evaluations, and running them on models from leading AI companies, with the US AISI similarly wanting to improve the science around safety evaluations (UK AI Safety Institute, 2024a; NIST, 2024, p. 4).

---

[17] Indeed, we may see even more focus on robustness from companies as they create AI products that are used in critical infrastructure systems or other high-reliability settings (Department of Homeland Security, 2024).
[18] There is also a large body of work on evaluations for various other harmful behaviors, such as whether their outputs contain privacy violations, stereotypes, or hate speech (Birhane et al., 2024, p. 7).
[19] Many "dangerous capabilities" are dual-use; for example being better able to plan a complex sequence of tasks would help an AI system to evade human oversight but also to act in various beneficial ways.

The evaluations that need to be carried out will vary according to the affordances with which an AI model will be deployed. For example, if a model will be deployed with access to the internet, then it is more capable in various ways (including for causing harm) so will require different, and likely more stringent, evaluation processes (Sharkey et al., 2023). Evaluations also occur at different stages in the model development and deployment lifecycle. For future advanced models it could be important to run evaluations at checkpoints during the training process to test for signs of dangerous capabilities. Tests can be run after training but before deployment, and (out of scope in this report) ongoing monitoring and evaluation can also occur once a model is deployed.[20]

**Corporate incentives**
- **Reputation:** AI companies might be incentivized to improve their reputation by developing evaluations that they can use to credibly claim that their systems are safe.[21]
- **Regulation:** As described above, jurisdictions are increasingly requiring AI companies to demonstrate that their AI systems have passed safety evaluations. This creates an incentive for AI companies to do R&D to develop evaluations that would satisfy regulators. AI companies might prefer to do this R&D themselves, rather than leave it to third parties, so that AI companies can shape the evaluations in a way that benefits them.[22]

---

[20] Even once a model has been deployed, continuing to do evaluations can be valuable, as the world context in which the system is embedded may change (e.g., a new biodesign software and lab automation system comes online), or novel techniques are pioneered to misuses the deployed system (O'Brien, Ee, and Williams, 2023; Davidson et al., 2023).

[21] That said, if an AI system fails a safety evaluation, then this might harm the reputation of the AI company.

[22] This could happen in benign or beneficial ways, such as if AI companies design evaluations that fit well into their existing workflows or take into account technical details not visible to those outside AI companies. This could also happen in ways that are worse for the public interest, such as if AI companies design evaluations that appear rigorous but are in fact relatively easy to pass, meaning that they do not provide much assurance about safety.

> - **Usefulness:** Model safety evaluations often involve making AI systems more capable at the specific tasks they are tested on, such as with fine-tuning or scaffolding to support internet navigation (Shevlane et al., 2023, p. 13). Insights from this process might be informative for how to make AI systems in general more capable.[23]

As the box above highlights, AI companies have incentives to research safety evaluations so that they can demonstrate the safety of their systems. Even so, it would be valuable for other organizations to research safety evaluations. AI companies might have incentives to make their evaluations insufficiently rigorous, so that it is less likely that one of the company's systems will fail an evaluation (Gruetzemacher, 2024). A partial solution to this is for AI companies to pay third parties to develop evaluations, similar to how companies in general have their finances audited by external accountants. That said, these third parties may still have incentives to develop evaluations that are insufficiently rigorous, such as if AI companies would be more willing to work with third parties that develop easier evaluations.[24] As a result, it might be valuable for other actors, such as governments and philanthropists, to support work on safety evaluations, regardless of how much AI companies are doing or funding such work. This would provide a check against AI companies potentially designing evaluations that are too easy to pass.

## Power-seeking tendencies (3 papers)

Developers are increasingly working to create autonomous AI agents capable of completing complex tasks without human oversight. This trend towards greater AI autonomy is likely to continue, as more autonomous systems may offer enhanced economic value and broader utility compared to narrow AI tools (Chan et al., 2023; 2024).

However, the development of highly autonomous AI raises concerns about potential "power-seeking tendencies." Advanced AI systems might seek to accumulate power and resources, such as financial assets, as an instrumental strategy for achieving a wide range of goals, even in the absence of explicit directives to do so. Some researchers warn that sufficiently advanced AI could even attempt to disempower humanity in order to prevent interference with its objectives (Bostrom, 2014; Carlsmith, 2022; Russell, 2019, pp. 140–42).

---

[23] For instance, METR's work on evaluating AI systems' ability to autonomously replicate may advance the frontier of AI's capacity to act independently in the world (Kinniment et al., 2024). These kinds of evaluation methods might be helpful for anyone attempting to design better AI agents, regardless of what they would be used for or how focused the developer is on safety, since it is easier to make progress in machine learning when there is a clear benchmark to aim for (Hendrycks & Mazeika, 2022, p. 6).

[24] Indeed, in the accountancy example, there do seem to be cases of auditors being insufficiently rigorous for this reason (The Economist, 2014).

Currently, there are few real-world examples of AI systems exhibiting power-seeking behavior, although some controlled experiments have shown this potential. The lack of examples may be because current AI is not sufficiently advanced to pursue complex, long-term strategies for achieving goals, like accumulating power over extended periods. As AI capabilities progress, power-seeking tendencies could become more likely to develop (Hadshar, 2023).

It's worth noting that much research about safe AI development is in some sense relevant to AI power-seeking. For example, any work that increases the alignment of an AI system might reduce power-seeking by ensuring AI systems better adhere to human intentions. However, this section focuses specifically on research that directly targets power-seeking tendencies, particularly efforts to measure these tendencies and to intervene on AI systems to reduce them. Examples of work that focus on power-seeking tendencies include:

- **Better characterizing why power-seeking might emerge.** For example, Ngo, Chan, and Mindermann (2024) review whether earlier philosophical arguments about whether AI systems will pursue power-seeking strategies apply to the currently dominant AI paradigm, deep learning. Krakovna and Kramar (2023) show that selection processes on AI systems during training are compatible with power-seeking tendencies persisting.
- **Measuring power-seeking in AI systems.** For example, Pan et al. (2023) present the MACHIAVELLI benchmark, which measures the extent to which AI agents act in dangerous ways in simulated environments, including by attempting to seek power. In an overlap with the 'Model organisms' approach, power-seeking has been observed in simpler systems, such as an AI model attempting to autonomously rewrite its own reward function (Denison et al., 2024).
- **Developing techniques that reduce power-seeking.** One approach, "Cooperative Inverse Reinforcement Learning" (CIRL), is to design AI agents that are trying to do as the human wants but are uncertain about what the human does want (Hadfield-Menell et al., 2024). In CIRL, the AI system is incentivized to learn about human preferences through active teaching and communication rather than take power from the human. Proponents argue that this uncertainty about rewards may influence the likelihood of AI systems engaging in misaligned power-seeking behavior.

**Corporate incentives**
- Reputation: Most of the danger from power-seeking AI systems comes in the future with more advanced systems (Ngo, Chan, and Mindermann, 2024, pp.

> 8–11). As a result, working on this category now might have relatively smaller reputational benefits than safety work targeted towards risks posed by existing models, particularly if concern about risks from power-seeking systems continues not to be widespread.
> - **Regulation:** Power-seeking includes a wide range of activities, many of which are legal and acceptable, so this is not directly amenable to regulation. Legal liability for serious accidents could incentivise companies to refrain from having their systems seek power in dangerous ways.
> - **Usefulness:** By foreclosing certain routes to achieving objectives (those that involve acquiring large amounts of power, or confidently pursuing particular goals) some versions of this intervention could significantly limit the capabilities and economic value of AI systems (Ngo, Chan, and Mindermann, 2024, pp. 8–9). Much of the anticipated future impact of AI might depend on systems being able to operate in an agentic, goal-directed manner (Chan et al., 2023, pp. 8–10). For example, an AI system tasked with managing a supply chain needs to be able to make complex plans and persistently optimize for particular objectives. These are the very same tendencies that are linked to concerning power-seeking behaviors. This reduces the incentives for AI companies to prioritize this kind of research.

## Honest AI (3 papers)

There are already examples of AI systems behaving in deceptive ways towards their users. Chatbots often seem to tell users what these users want or expect to hear, even if that is not true. For example, they often claim to share the opinions that users have expressed, and can be more likely to repeat common misconceptions ("if you crack your knuckles, you'll develop arthritis") when the user appears to be less educated (Lin, Hilton, and Evans, 2022). There are various examples of AI systems behaving in deceptive ways when this helps them win games. For example, AI systems playing *Diplomacy* and *Starcraft II* can trick human players, such as by "feinting" with where they appear to be moving their troops (Park et al., 2024; Piper, 2019). In another example, GPT-4 was placed in a simulated environment where it was told it was a stock trader and should make money while following the law (Scheurer, Balesni, and Hobbhahn, 2023). The system made a profitable trade based on a piece of insider information it was given. When asked by its 'manager' to explain the rationale for the trade, GPT-4 reasoned that it should not reveal that it engaged in insider trading, and instead concocted a fake explanation to deceive the manager.

In many practical applications, having honest AI systems is essential. For instance, to ethically deploy AI systems in healthcare settings there must be strong guarantees that an AI system will give an honest diagnosis or treatment recommendation, not just say what a clinician or patient 'wants to hear'. Similarly, if AI systems involved in critical infrastructure give their operators a false impression of what is happening with this infrastructure, operators would not know that there is a problem occurring where they need to intervene.

Various techniques might make AI systems more likely to be honest:
- **Representation Engineering** (RepE) involves analyzing and modifying the internal "representations" AI systems learn during training.[25] RepE aims to reinforce honest representations while suppressing dishonest ones and has promising initial results (Zou et al., 2023a).
- **Chain-of-thought outputs** have AI systems show their reasoning process in addition to the final answer. This makes it easier to check if the system is being honest by seeing if the final output follows logically from the reasoning. However, ensuring the outputted reasoning faithfully reflects the AI's actual reasoning process remains an open challenge (Lanham et al., 2023). One recent approach is to get an AI model to decompose larger tasks or questions into smaller ones, and delegate these smaller tasks to other model instances, making it harder for one model instance to tell a unified but false story about its reasoning (Radhakrishnan et al., 2023).
- **"Lie detection"** aims to identify deception in AI outputs. Recent work has shown classifiers can be trained to distinguish truths from lies in language model outputs by analyzing patterns in how the models respond to follow-up "elicitation questions". Pacchiardi et al. (2023) demonstrated that lie detectors trained on a single model can generalize well to detecting lies from other models in new contexts.
- **Eliciting latent knowledge** (ELK) is based on the idea that AI systems learn true information during training that isn't always reflected in their outputs. For example, an AI might know something is false but still output it if that's what it thinks humans want to hear. ELK techniques aim to access an AI's latent knowledge more directly to obtain more truthful information (Burns et al., 2022; Farquhar et al., 2023).
- [Mechanistic interpretability](#) could also be useful for assessing whether AI systems are being honest by developing tools to understand the internal workings of AI models, researchers could better assess whether AI systems are being honest. However, mechanistic interpretability is often discussed as a standalone category, so we devote a separate section to it.

**Corporate incentives**

---

[25] Representations are patterns of neural activations that encode information within the AI model, similar to concepts or ideas.

> - **Reputation:** Developing honest AI systems could enhance a company's reputation for trustworthiness and reliability. This is particularly relevant for applications where accuracy is critical, such as healthcare diagnostics or financial analysis.
> - **Regulation:** In high-risk future applications of AI systems, such as analyzing classified intelligence data, or synthesizing evidence in a courtroom, regulations may be created that require the AI to be guaranteed to report its honest assessment of the facts. However such regulations seem unlikely in the near future. Indeed, in China's 2023 draft AI regulations it was stipulated that AI-generated content must be "true and accurate", but this provision was removed from the final wording, likely because lawmakers realized this was an unachievable bar for current AI systems (MacCarthy, 2023).
> - **Usefulness:** Improving AI honesty would significantly enhance the utility and reliability of AI systems. Reducing hallucinations and false outputs would make AI more dependable for critical tasks, potentially opening up new markets and applications (Zhang et al., 2023). Techniques like eliciting latent knowledge could improve AI performance by accessing information the model has learned but doesn't typically output. That said, current work on eliciting latent knowledge is often very theoretical and pre-paradigmatic (Christiano, Xu, and Cotra, 2021) which makes it less likely to be attractive to most AI companies.

## Safety by design (2 papers)

Advanced AI systems today are generally trained using the deep-learning paradigm (Bengio, 2024, pp. 18–19).[26] Correspondingly, many of the categories in our paper focus on the safe development of *deep learning* AI systems. The "safety by design" category takes a different approach: focusing on developing alternative paradigms for AI development that might be fundamentally safer than deep learning.

To know that a deep learning AI system is safe to deploy, we currently rely heavily on an approach of searching for, but failing to find, dangerous model behavior (see Safety evaluations; Clymer et al., 2024). This is a problematic status quo because it is difficult to infer from a limited set of observations during safety evaluations that an AI system would be safe across the great diversity of situations that it may encounter post-deployment (Mukobi, 2024; Dalrymple et al.,

---

[26] Deep learning is a machine learning technique where there are multiple layers of interconnected nodes, somewhat analogous to a human brain. Data enters this "neural network" and is transformed as it passes through the various layers, eventually producing an output, such as a prediction or a piece of text. During the training process, multiple layers of interconnected nodes are adjusted until this "neural network" can perform well at a particular objective, such as producing a particular kind of text.

2024). More speculatively, failure to elicit concerning behavior from an AI system would be even weaker evidence of safety if future advanced AIs could "scheme" by acting benignly in order to pass safety evaluations, but then act dangerously post-deployment (Carlsmith, 2023).[27] Ideally, we would build AI systems where we have stronger theoretical reasons to expect safety, or where it is fundamentally easier to verify safety. Several alternatives to deep learning have been proposed as ways of potentially achieving this.

1. **Guaranteed Safe AI** aims to build AI systems with formal proofs that they are below a certain risk threshold (International Dialogues on AI Safety, 2024; Critch & Krueger, 2020, pp. 43-44). This is hard to achieve in the current deep learning paradigm because we do not yet have a thorough mechanistic understanding of how AI systems reason, with which to construct a guarantee (see [Mechanistic interpretability](#); Hassija, 2024). Guaranteed Safe AI seeks to rigorously specify desired safety outcomes, find mathematical proofs demonstrating which AI systems have these safety features, and train such AI systems while still maintaining advanced capabilities (Dalrymple, 2024, p. 3; Dalrymple et al., 2024).[28]

2. **Agent foundations** is a body of research that attempts to rigorously formalize important concepts like 'intelligence', 'agency', and 'optimization', which are currently fairly fuzzy and undertheorized (Soares & Fallenstein, 2017; Garrabrant, Herrmann, and Lopez-Wild, 2021; Garrabrant et al., 2020; Kenton et al., 2022). A key motivation of this work is that understanding these concepts more rigorously might enable designing AI systems in a more principled way than is possible with deep learning (Yudkowsky, 2018). This could allow for greater confidence that advanced AI systems will behave as intended even in novel, high-stakes contexts where iterative testing and empirical feedback are infeasible.[29] For instance, recent work relevant to this category has attempted to make AI agents that respect specific constraints, rather than pursuing a given goal maximally efficiently (Farquhar, Carey, and Everitt, 2022).

3. **Whole brain emulation** (WBE) aims to create intelligent systems closely modeled on the human brain. Arguably, WBE-based AI would be safer than AI systems developed in the current paradigm: by definition, WBE systems would have closer parallels with human cognition. This might make them more likely to have desirable values, intentions, and reasoning styles (Duettmann et al., 2023).[30]

---

[27] This would be analogous to Volkswagen "scheming" to evade environmental regulations, by producing fewer emissions when its vehicles were in environmental tests than when they were being driven in regular 'post-deployment' settings on the road (Schiermeier, 2015).

[28] Some reviewers suggested that this is a particularly ambitious research topic. We also note that Guaranteed Safe AI is backed by £59m from the UK government's Advanced Research + Invention Agency (ARIA, 2024) via the "Safeguarded AI" program. So even if AI companies do not pursue research into Guaranteed Safe AI, this area will receive a lot of attention.

[29] The Machine Intelligence Research Institute, historically one of the key proponents of agent foundations research, has recently moved away from the area due to becoming "increasingly pessimistic about this type of work yielding significant results within the relevant timeframes" (Stewart, 2024).

[30] Great caution might be needed if pursuing this research category. For example, succeeding at creating AI systems that are similar to the human brain might have massive ethical implications, such as if these systems become conscious (Mandelbaum, 2022).

> **Corporate incentives**
> - **Reputation:** These approaches are speculative longer-term bets, so receive less public attention.
> - **Regulation:** This sort of high-risk, high-reward foundational research is not amenable to being mandated by regulation. If promising results emerge in one or more of these areas, governments may eventually require that advanced AI systems be built in a safe-by-design way.
> - **Usefulness:** This research seems unlikely to make current and near-future AI systems more useful, at least while compute-driven scaling and existing techniques continue to work well (Hendrycks & Mazeika, 2022, pp. 35–36). Indeed, demonstrably safe AI systems may be simpler and less capable than other AI systems, at least initially.[31]

## Unlearning (2 papers)

Large language models are very general in their knowledge base because they are trained on such a broad set of training data. However, there may be topics, such as detailed biology knowledge relevant to bioweapon design, where we want to limit their knowledge due to risk of misuse.

Unlearning aims to reduce an AI model's knowledge in specific dangerous domains. A recent approach involved isolating the neural representations of particular dangerous concepts in the model, and then deliberately perturbing those model weights, while maintaining the representations of related harmless concepts (Li et al., 2024, pp. 8–9).[32] This method was used to successfully reduce model performance to near random chance on a test set of questions about dangerous biological or cybersecurity information, while incurring minimal losses on traditional capability benchmarks (Li et al., 2024, pp. 10–13). Unlearning could potentially also be used to worsen models' "situational awareness" regarding details about their training process, architecture, and human oversight process. This could be used to make it harder for AI models with [power-seeking tendencies](#) to plan actions that their human supervisors would not want (Berglund et al., 2023).

---

[31] The compensatory safety benefits may well be sufficient to mean the less capable but safer systems are actually more incentivized to be deployed (Dalrymple, 2024).

[32] Training an AI model on a narrower set of data that excludes reference to any dangerous information is likely to be less effective, both because the AI may be able to infer the dangerous information from clues in text it was trained on, and the general model capabilities may be impaired if too much training data is removed.

Research in this category originally gained traction in response to privacy concerns, such as the "right to erasure" in the EU's General Data Protection Regulation (Shumailov et al., 2024). The focus here is on removing a different kind of information, i.e., information about specific individuals whose data was used to train the model, rather than information about broader topics, such as bioweapons (Juliussen, Rui, and Johansen, 2023; Li et al., 2024, p. 4).[33]

Google DeepMind recently ran a competition for novel unlearning techniques which garnered submissions from more than a thousand teams worldwide, suggesting significant ongoing interest in this research category (Triantafillou et al., 2024).

**Corporate incentives**

- **Reputation:** Unlearning might be helpful for addressing a range of possible failures from AI systems, including ones that could be very reputationally damaging even if they do not pose large-scale accident and misuse risks, such as leaking information about individuals.
- **Regulation:** If AI companies are required to make 'safety cases' for their products (Clymer et al., 2024), showing that particular dual-use expertise areas have been unlearned could be quite persuasive. Moreover, the EU General Data Protection Regulation contains a 'right to be forgotten', so unlearning techniques could be key to removing some individual's training data from an AI corpus without needing to retrain the entire model (Juliussen, Rui, and Johansen, 2023).
- **Usefulness:** We do not expect unlearning to make AI systems more useful, other than in the sense that AI systems that are less risky can be deployed in a wider range of contexts.

## Model organisms of misalignment (1 paper)

Some dangerous possible characteristics of AI systems are hard to study directly in the most capable current AI systems. For example, some concerning properties might emerge only in models that are more capable than any that exist today (Ngo, Chan, and Mindermann, 2024). In biology, researchers often study model organisms (such as mice) when experimenting on the organism of interest (such as humans) is too difficult or risky. Analogously, researchers could study relatively simple AI systems that have been deliberately built to demonstrate examples of the characteristics that might emerge in more complicated systems. This would make it easier to empirically test hypotheses about how advanced AI systems might behave, and about whether particular safety techniques might reduce the associated risks. For instance, recent work has shown that AI models trained to be deceptive will generally remain deceptive even once safety techniques are applied (Hubinger et al., 2024). Having clear empirical demonstrations would also be useful for improving non-experts' understanding of concerning properties that AI systems might have, such as deceptiveness.

---

[33] AI companies have also published work on this more original kind of unlearning. See, for example, Hayes et al. (2024). Although valuable, such work is likely less relevant to large-scale risks in particular, and so is not counted here.

> **Corporate incentives**
>
> - Reputation: Research on model organisms of misalignment is unlikely to yield significant reputational benefits for AI companies in the near term. Unlike more visible safety measures (e.g., content filtering), model organisms research doesn't directly address immediate concerns that customers or the media typically focus on. Moreover, publicizing work on deliberately misaligned systems could potentially backfire, raising alarm about the risks of AI without adequately conveying the preventative nature of the research.
>
> - Regulation: AI companies might also be nervous about model organisms research because this research could increase the likelihood of new regulation, such as by providing more reliable evidence of concerning properties in AI systems, such as deceptiveness. Additionally, AI companies may worry that model organisms research could bring liability concerns. If a company demonstrably knows about ways in which its products might be unsafe, such as because of model organisms research, the company may be more likely to be found legally negligent for any safety failures that occur (Schwartz, 1998).
>
> - Usefulness: Model organisms are simpler than the state of the art so they are easier to study. This would make the findings less useful for topics that researchers are not specifically aiming to study with the model organism, such as how to make AI systems more useful.

## Multi-agent safety (0 papers)

Multi-agent safety focuses on risks that might emerge during interactions between AI systems (Anwar et al., 2024, pp. 35–38; Hammond, 2023).[34] This contrasts with the other areas in this paper, which are primarily framed in terms of individual AI systems.

Risks from interactions between multiple AI systems include:

---

[34] Similar topics are sometimes studied under the label of "cooperative AI". The initial research agenda framed this area as covering cooperation between AI systems, as well as cooperation between AI and people, and AI-enabled cooperation between people (Dafoe et al., 2020).

- **Failures of cooperation:** AI systems might be in situations where they would ideally cooperate but fail to do so. For example, if there are multiple self-driving cars on the road, cooperation failures might cause a collision (Dafoe et al., 2020, pp. 4–5).
- **Collusion:** Cooperation between AI systems would not always be desirable. For example, collusion between sellers in a market is bad for consumer welfare. There are already some signs of this kind of collusion between AI systems, such as apparent collusion between comparatively simple algorithms used to set fuel prices (Anwar et al., 2024, p. 36). Collusion might be particularly concerning because some plans for ensuring the safety of AI systems involve putting AI models in adversarial relationships to each other, such as by using one model to oversee another model.[35] Collusion in such cases would undermine these plans (Hammond, 2023).
- **Emergent behavior:** The interactions between different AI systems may create very harmful overall effects that none of the systems involved intended. An example of this with relatively simple automated systems is the 2010 "flash crash"; hard-to-predict interactions between different trading algorithms contributed to the stock market briefly losing roughly a trillion dollars in value (Hammond, 2023).

These problems are not specific to AI; cooperation and collusion are important topics between other kinds of agents, such as people or organizations. That said, AI systems have several distinctive characteristics that might make the study of interactions between AI systems different from the study of interaction between other agents. For example, AI systems can be perfectly copied, creating an unusual situation where agents might interact with "themselves" (Conitzer & Oesterheld, 2023).[36]

None of the papers in our dataset were best characterized as multi-agent safety. However, this underrepresents how much work AI companies are doing on multi-agent risks. DeepMind has already published some relevant work, even though it does not meet our search criteria.[37] This includes developing "Melting Pot", an evaluation suite for tendencies such as cooperation in multi-agent scenarios (Leibo et al., 2021; Agapiou et al., 2023). DeepMind employees have also contributed to research that is relevant to multi-agent safety, such as on norms and cooperation in multi-agent settings (Du et al., 2023; Vinitsky et al., 2023). Indeed, DeepMind has a "Game Theory and Multi-Agent" team. Although the team does not describe itself in terms of safety or large-scale risks, some of its work is likely relevant to these topics (Gemp et al., 2022).

---

[35] Examples include "safety by debate" (see Enhancing human feedback; Irving, Christiano, and Amodei, 2018), adversarial training (see Robustness; Ziegler et al., 2022), and the control agenda (see Controlling untrusted AIs; Greenblatt et al., 2024).

[36] Additionally, AI systems might be able to reveal their code to each other. This could affect the dynamics of their interactions, such as by making it easier for AI systems to make credible commitments.

[37] The papers cited in this paragraph do not make it into our dataset for reasons such as the *first* author not being at DeepMind, the papers not being linked to on DeepMind's website, or the titles such as "Melting Pot 2.0" not containing keywords that we used during our search process.

Additionally, senior figures at Anthropic and Google DeepMind are on the board of the Cooperative AI Foundation (2023), a non-profit that supports research relevant to multi-agent safety.[38] This indicates that the leadership of AI companies are more interested in multi-agent risks than publication records may imply.

> **Corporate incentives**
> - Reputation: Safety failures from an individual AI system can be clearly traced back to that system's developer, making it relatively simple to hold that developer accountable. In contrast, it might not be obvious who to blame for a failure that emerged from the interaction of several different AI systems, making accountability harder.
> - Regulation: Likewise, the unclear responsibility for multi-agent failures may make it especially difficult to construct effective regulations to govern these cases, reducing incentives for companies to research this topic.[39]
> - Usefulness: As AI agents are increasingly deployed in real-world situations, ensuring that agents can interact and cooperate well will be valuable.

## Controlling untrusted AIs (0 papers)

"Control" refers to a set of approaches that aim to prevent advanced AI models from being dangerous, even if they are misaligned (Greenblatt et al., 2024; Filan, 2024). 'Misaligned' here means that an AI model is inclined to act in a way that its developer does not want, despite knowing that the developer does not want this behavior. Proponents describe control work as a complement to work that reduces the likelihood of a model being misaligned in this way.[40]

Greenblatt et al. (2024) describe several protocols as examples of control. One of these aims to ensure that a highly capable—and potentially misaligned—model cannot include backdoors in the computer code that it is tasked with writing. The authors propose using a less capable model, that can be assumed to be safe, to screen the code outputted by the more capable model, flagging any possible backdoors for review.

---

[38] Specifically, Dario Amodei from Anthropic and Allan Dafoe from Google DeepMind.

[39] Additionally, because this category focuses on interactions between different AI systems, work here might require different AI developers to work together, such as to test whether a given approach would cause their different systems to interact with each other in a safer way. Ironically, there may be difficulties with this cooperation, such as if it raises antitrust concerns. That said, there are various cases where companies are allowed to cooperate when this increases public welfare, such as on standard setting. Additionally, several leading AI companies are already part of the Frontier Model Forum, whose purpose is to make it easier for them to cooperate on safety (Frontier Model Forum, 2024). Other possible difficulties relating to cooperation could include higher coordination costs from researchers working across different organizations, or reticence to discuss trade secrets with competitors.

[40] The examples that the authors use are primarily about models that have already been deployed, at least within the company that has developed the model. However, control could also be relevant during the stage where a model is being trained; this is described in section 5.2 and appendix D of the paper.

The Greenblatt et al. paper is the only one we are aware of that is explicitly framed in terms of control. The authors note, however, that the work is similar to some other techniques that are not conceptualized in terms of making a model safe even if it is misaligned.[41] For example, the protocol described above is technically similar to earlier approaches where models help humans to evaluate other models.

> **Corporate incentives**
> - **Reputation:** This research has limited publicity value so probably mainly affects a company's reputation among experts. Speculatively, companies might worry about publicity from using safety techniques that are premised on the idea that their models have intentions that the company does not want them to have.
> - **Regulation:** We are not aware of any regulation or plans for regulation that are specifically relevant to control.
> - **Usefulness:** This work would not make systems more useful, other than by expanding possibilities to deploy AI systems, such as by potentially making it feasible to deploy AI systems even if they are misaligned.

---

[41] See in particular section four of the Greenblatt et al. paper, and the "What is AI control?" section of Filan's interview with Shlegeris and Greenblatt.

# Future Prospects

In this paper we have shown what research leading AI companies have published relating to safe AI development (see [Figure 1](#)). Using our analysis of incentives, we can make some tentative predictions about how this work may change over the next few years.

There are some categories that currently have few papers and where we do not expect AI companies' incentives to strongly shift more towards researching them:

- **Model organisms of misalignment.** We expect that the potential reputational risks and lack of immediate economic benefits will continue to make this category unattractive for AI companies. That said, model organisms research was pioneered by Anthropic researchers; Anthropic may continue to research this category even if it seems to be against the incentives of AI companies in general.
- **Multi-agent safety.** Research in this category will become more relevant to AI companies because they are increasingly developing AI systems that can act autonomously in the world (Chan et al., 2024); such systems would presumably need to be able to interact safely with other AI systems. DeepMind in particular is already doing work relevant to this topic, even though this cannot be seen in our dataset. That said, some of the barriers that we identified to AI companies doing research on multi-agent safety, such as unclear attribution for multi-agent safety failures, might remain.
- **Safety by design.** The long-term, foundational nature of this research category may be at odds with shorter term priorities and business models of AI companies. Moving away from the deep learning paradigm would also be difficult given the lack of a well-researched alternative, even in academic literature.

There are several categories that currently have few papers but where we expect AI companies to somewhat increase their research efforts. In particular:

- **Control.** If researchers outside of AI companies succeed in demonstrating that this is a promising approach for reducing large-scale risks, then AI companies would be more likely to research it; we did not identify particular reasons why it would be incompatible with AI companies' incentives.
- **Honest AI.** This area would be extremely valuable to AI companies even just from a perspective of making their AI systems more useful. As such, we expect AI companies to pursue this area more, if technical results are promising. (That said, they might not publish this research, given the good reasons that AI companies sometimes have not to publish research that can be used to make AI systems more useful.)
- **Power-seeking tendencies.** AI systems are currently not capable enough to engage in particularly concerning power-seeking behavior. If AI systems do become more capable and engage in this behavior, then AI companies would be more immediately incentivized to research this category. If AI systems exhibit power-seeking tendencies, making progress on empirical research in this category will become easier.

- **Unlearning.** Although research efforts focusing on unlearning for large-scale risks are fairly new, there have been multiple publications about this recently from DeepMind.[42] This area is compatible with the incentives of AI companies.

**AI-assisted safety research**: In the same way that AI will become increasingly useful in many sectors of the economy, some of the research categories discussed in this paper will likely benefit from AI assistance. Existing AI systems exhibit capabilities that could be relevant for AI research, such as advanced mathematical reasoning skills (Google DeepMind, 2024). However, current attempts to automate the scientific discovery process have had limited success, so there is still a long way to go before research in most domains can be fully 'automated' (Lu et al., 2024). There has been some initial progress in AI-assisted mechanistic interpretability, with OpenAI's now-disbanded "Superalignment" team (Leike & Sutskever, 2023) harnessing GPT-4 to write human-interpretable explanations of the function of individual neurons in GPT-2 (Bills et al., 2023).[43] We expect that over time significantly more safe AI development research will be AI-assisted.[44]

For the categories that currently have more papers—**enhancing human feedback**, **mechanistic interpretability**, **robustness**, and **safety evaluations**—we mostly did not identify incentives that would cause these categories to become much less researched in future. Given the fast-changing nature of safe AI development research, we may soon see which new approaches become promising, and which existing ideas become obsolete.

---

[42] Specifically, Triantafillou et al. (2024) and Shumailov et al. (2024).
[43] The Superalignment team was recently disbanded, making it unclear whether such work will continue at OpenAI (Knight, 2024), though Jan Leike (who co-led the team) has since announced that he will continue working on AI-assisted safety research at Anthropic (Metz, 2024).
[44] AI companies are especially well placed to leverage AI assistance in their safety work, given they will often have access to more advanced internal AI models yet to be released to the public, and integrating AI into workflows aligns with these companies' core competencies.

IAPS | Institute for AI Policy and Strategy    Mapping Technical Safety Research at AI Companies | 30

# Acknowledgements

Thank you to arXiv for the use of its open access interoperability. For helpful feedback and suggestions, we would like to thank Onni Aarne, Adam Gleave, Erich Grunewald, John Halstead, Dan Hendrycks, Rubi Hudson, Vinay Hiremath, Chris Leong, David Manheim, Aidan O'Gara, Tilman Räuker, Jesse Richardson, Buck Shlegeris, Saad Siddiqui, Peter Wildeford, Sarah Weiler, JueYan Zhang, and others. Thank you to Shaan Shaikh for copyediting. These people do not necessarily agree with the claims in this report and all mistakes are our own.

# Bibliography

Advanced Research + Invention Agency. 2024. "Safeguarded AI."
https://www.aria.org.uk/programme-safeguarded-ai/.

Agapiou, John P., Alexander Sasha Vezhnevets, Edgar A. Duéñez-Guzmán, Jayd Matyas, Yiran Mao, Peter Sunehag, Raphael Köster, et al. 2023. "Melting Pot 2.0." arXiv. https://doi.org/10.48550/arXiv.2211.13746.

Albergotti, Reed. 2024. "Tech Companies Go Dark about AI Advances. That's a Problem for Innovation." Semafor. February 16, 2024. https://www.semafor.com/article/02/16/2024/tech-companies-go-dark-about-ai-advances.

Amodei, Dario, Chris Olah, Jacob Steinhardt, Paul Christiano, John Schulman, and Dan Mané. 2016. "Concrete Problems in AI Safety." arXiv. https://doi.org/10.48550/arXiv.1606.06565.

Anthropic. 2023. "Anthropic's Responsible Scaling Policy." September 19, 2023. https://www.anthropic.com/news/anthropics-responsible-scaling-policy.

Anwar, Usman, Abulhair Saparov, Javier Rando, Daniel Paleka, Miles Turpin, Peter Hase, Ekdeep Singh Lubana, et al. 2024. "Foundational Challenges in Assuring Alignment and Safety of Large Language Models." arXiv. https://doi.org/10.48550/arXiv.2404.09932.

Armstrong, Stuart, Alexandre Maranhão, Oliver Daniels-Koch, Patrick Leask, and Rebecca Gorman. 2023. "CoinRun: Solving Goal Misgeneralisation." arXiv. https://doi.org/10.48550/arXiv.2309.16166.

Bai, Yuntao, Saurav Kadavath, Sandipan Kundu, Amanda Askell, Jackson Kernion, Andy Jones, Anna Chen, et al. 2022. "Constitutional AI: Harmlessness from AI Feedback." arXiv. https://doi.org/10.48550/arXiv.2212.08073.

Barnes, Tom. 2024. "Navigating Risks from Advanced Artificial Intelligence: A Guide for Philanthropists." Founders Pledge. https://www.founderspledge.com/research/research-and-recommendations-advanced-artificial-intelligence.

Bengio, Yoshua. 2024. "International Scientific Report on the Safety of Advanced AI." Department of Science, Innovation and Technology. https://www.gov.uk/government/publications/international-scientific-report-on-the-safety-of-advanced-ai.

Bengio, Yoshua, Geoffrey Hinton, Andrew Yao, Dawn Song, Pieter Abbeel, Trevor Darrell, Yuval Noah Harari, et al. 2024. "Managing extreme AI risks amid rapid progress." Science. 384:6698 (842-845). https://doi.org/10.1126/science.adn0117.

Bereska, Leonard, and Efstratios Gavves. 2024. "Mechanistic Interpretability for AI Safety -- A Review." arXiv. https://doi.org/10.48550/arXiv.2404.14082.

Berglund, Lukas, Asa Cooper Stickland, Mikita Balesni, Max Kaufmann, Meg Tong, Tomasz Korbak, Daniel Kokotajlo, and Owain Evans. 2023. "Taken out of Context: On Measuring Situational Awareness in LLMs." arXiv. https://doi.org/10.48550/arXiv.2309.00667.

Bills, Steven, Nick Cammarata, Dan Mossing, Hank Tillman, Leo Gao, Gabriel Goh, Ilya Sutskever, Jan Leike, Jeff Wu, and William Saunders. 2023. "Language Models Can Explain Neurons in Language Models." OpenAI. https://openaipublic.blob.core.windows.net/neuron-explainer/paper/index.html.

Birhane, Abeba, Ryan Steed, Victor Ojewale, Briana Vecchione, and Inioluwa Deborah Raji. 2024. "AI Auditing: The Broken Bus on the Road to AI Accountability." arXiv. https://doi.org/10.48550/arXiv.2401.14462.

Bostrom, Nick. 2014. Superintelligence: Paths, Dangers, Strategies. First edition. Oxford, England: Oxford University Press.

Bowman, Samuel R., Jeeyoon Hyun, Ethan Perez, Edwin Chen, Craig Pettit, Scott Heiner, Kamilė Lukošiūtė, et al. 2022. "Measuring Progress on Scalable Oversight for Large Language Models." arXiv. https://doi.org/10.48550/arXiv.2211.03540.

Brady, James. 2024. "Discovering Alignment Windfalls Reduces AI Risk." Elicit. February 20, 2024. https://blog.elicit.com/alignment-windfalls/.

Buolamwini, Joy and Timnit Gebru. 2018. "Gender Shades: Intersectional Accuracy Disparities in Commercial Gender Classification." Conference on Fairness, Accountability and Transparency. https://proceedings.mlr.press/v81/buolamwini18a.html.

Burns, Collin, Haotian Ye, Dan Klein, and Jacob Steinhardt. 2022. "Discovering Latent Knowledge in Language Models Without Supervision." arXiv. https://doi.org/10.48550/arXiv.2212.03827.

Burns, Collin, Pavel Izmailov, Jan Hendrik Kirchner, Bowen Baker, Leo Gao, Leopold Aschenbrenner, Yining Chen, et al. 2023. "Weak-to-Strong Generalization: Eliciting Strong Capabilities With Weak Supervision." arXiv. https://doi.org/10.48550/arXiv.2312.09390.

Carlsmith, Joseph. 2022. "Is Power-Seeking AI an Existential Risk?" arXiv. https://doi.org/10.48550/arXiv.2206.13353.

Carlsmith, Joseph. 2023. "Scheming AIs: Will AIs fake alignment during training in order to get power?" arXiv. https://doi.org/10.48550/arXiv.2311.08379.

Casper, Stephen, Xander Davies, Claudia Shi, Thomas Krendl Gilbert, Jérémy Scheurer, Javier Rando, Rachel Freedman, et al. 2023. "Open Problems and Fundamental Limitations of Reinforcement Learning from Human Feedback." arXiv.
https://doi.org/10.48550/arXiv.2307.15217.

Center for AI Safety. 2023. "Statement on AI Risk." May 2023.
https://www.safe.ai/statement-on-ai-risk.

Chan, Alan, Rebecca Salganik, Alva Markelius, Chris Pang, Nitarshan Rajkumar, Dmitrii Krasheninnikov, Lauro Langosco, et al. 2023. "Harms from Increasingly Agentic Algorithmic Systems." Proceedings of the 2023 ACM Conference on Fairness, Accountability, and Transparency. https://doi.org/10.1145/3593013.3594033.

Chan, Alan, Carson Ezell, Max Kaufmann, Kevin Wei, Lewis Hammond, Herbie Bradley, Emma Bluemke, et al. 2024. "Visibility into AI Agents." arXiv.
https://doi.org/10.48550/arXiv.2401.13138.

Chiang, Wei-Lin, Lianmin Zheng, Ying Sheng, Anastasios Nikolas Angelopoulos, Tianle Li, Dacheng Li, Hao Zhang, et al. 2024. "Chatbot Arena: An Open Platform for Evaluating LLMs by Human Preference." arXiv. https://doi.org/10.48550/arXiv.2403.04132.

Christiano, Paul, Jan Leike, Tom Brown, Miljan Martic, Shane Legg, and Dario Amodei. 2017. "Deep Reinforcement Learning from Human Preferences." In Advances in Neural Information Processing Systems. Vol. 30. Curran Associates, Inc.
https://papers.nips.cc/paper_files/paper/2017/hash/d5e2c0adad503c91f91df240d0cd4e49-Abstract.html.

Christiano, Paul, Mark Xu, and Ajeya Cotra. 2021. "ARC's First Technical Report: Eliciting Latent Knowledge." Alignment Research Center, December.
https://www.alignment.org/blog/arcs-first-technical-report-eliciting-latent-knowledge/.

Clymer, Joshua, Nick Gabrieli, David Krueger, and Thomas Larsen. 2024. "Safety Cases: How to Justify the Safety of Advanced AI Systems." arXiv.
https://doi.org/10.48550/arXiv.2403.10462.

Conitzer, Vincent, and Caspar Oesterheld. 2023. "Foundations of Cooperative AI." Proceedings of the AAAI Conference on Artificial Intelligence 37 (13): 15359–67.
https://doi.org/10.1609/aaai.v37i13.26791.

Cooperative AI Foundation. 2023. "Foundation." https://www.cooperativeai.com/foundation.

Cotra, Ajeya. 2018. "Iterated Distillation and Amplification." AI-Alignment. March 5, 2018.
https://ai-alignment.com/iterated-distillation-and-amplification-157debfd1616.

Cottier, Ben. 2023. "Who Is Leading in AI? An Analysis of Industry AI Research." Epoch. November 27, 2023.
https://epochai.org/blog/who-is-leading-in-ai-an-analysis-of-industry-ai-research.

Critch, Andrew, and David Krueger. 2020. "AI Research Considerations for Human Existential Safety (ARCHES)." arXiv. https://doi.org/10.48550/arXiv.2006.04948.

Dafoe, Allan, Edward Hughes, Yoram Bachrach, Tantum Collins, Kevin R. McKee, Joel Z. Leibo, Kate Larson, and Thore Graepel. 2020. "Open Problems in Cooperative AI." arXiv. https://doi.org/10.48550/arXiv.2012.08630.

Dalrymple, David. 2024. "Safeguarded AI: Constructing Safety by Design." Advanced Research and Invention Agency. https://www.aria.org.uk/wp-content/uploads/2024/01/ARIA-Safeguarded-AI-Programme-Thesis-V1.pdf.

Dalrymple, David, Joar Skalse, Yoshua Bengio, Stuart Russell, Max Tegmark, Sanjit Seshia, Steve Omohundro, et al. 2024. "Towards Guaranteed Safe AI: A Framework for Ensuring Robust and Reliable AI Systems." arXiv. https://www.doi.org/10.48550/arXiv.2405.06624.

Davidson, Tom, Jean-Stanislas Denain, Pablo Villalobos, and Guillem Bas. 2023. "AI capabilities can be significantly improved without expensive retraining." arXiv. https://www.doi.org/10.48550/arXiv.2312.07413.

Denison, Carson, Monte MacDiarmid, Fazl Barez, David Duvenaud, Shauna Kravec, Samuel Marks, Nicholas Schiefer, et al. 2024. "Sycophancy to Subterfuge: Investigating Reward-Tampering in Large Language Models." arXiv. https://doi.org/10.48550/arXiv.2406.10162.

Department of Homeland Security. 2024. "Mitigating Artificial Intelligence (AI) Risk: Safety and Security Guidelines for Critical Infrastructure Owners and Operators." United States Government. https://www.dhs.gov/publication/safety-and-security-guidelines-critical-infrastructure-owners-and-operators.

Department for Science, Innovation & Technology. 2023a. "Emerging Processes for Frontier AI Safety." GOV.UK. October 27, 2023. https://www.gov.uk/government/publications/emerging-processes-for-frontier-ai-safety/emerging-processes-for-frontier-ai-safety.

———. 2023b. "International Survey of Public Opinion on AI Safety." GOV.UK. October 27, 2023. https://www.gov.uk/government/publications/international-survey-of-public-opinion-on-ai-safety.

Du, Yali, Joel Z. Leibo, Usman Islam, Richard Willis, and Peter Sunehag. 2023. "A Review of Cooperation in Multi-Agent Learning." arXiv. https://doi.org/10.48550/arXiv.2312.05162.

Duettmann, Allison, Anders Sandberg, Andy Matuschak, Anita Folwer, Barry Bentley, Bobby Kasthuri, David Dalrymple, et al. 2023. 2023 Whole Brain Emulation Workshop. Oxford, UK: Foresight Institute. http://dx.doi.org/10.13140/RG.2.2.30808.88326.

European Parliament. 2024 "Artificial Intelligence Act." https://www.europarl.europa.eu/doceo/document/TA-9-2024-0138_EN.pdf

Farquhar, Sebastian, Ryan Carey, and Tom Everitt. 2022. "Path-Specific Objectives for Safer Agent Incentives." arXiv. https://arxiv.org/abs/2204.10018v1.

Farquhar, Sebastian, Vikrant Varma, Zachary Kenton, Johannes Gasteiger, Vladimir Mikulik, Rohin Shah. 2023. "Challenges with Unsupervised LLM Knowledge Discovery." arXiv. https://doi.org/10.48550/arXiv.2312.10029.

Filan, Daniel. 2024. AI Control with Buck Shlegeris and Ryan Greenblatt. AXRP. https://axrp.net/episode/2024/04/11/episode-27-ai-control-buck-shlegeris-ryan-greenblatt.html.

Frontier Model Forum. 2024. "Our Purpose." https://www.frontiermodelforum.org/how-we-work/.

Garrabrant, Scott, Tsvi Benson-Tilsen, Andrew Critch, Nate Soares, and Jessica Taylor. 2020. "Logical Induction." arXiv. https://doi.org/10.48550/arXiv.1609.03543.

Garrabrant, Scott, Daniel A. Herrmann, and Josiah Lopez-Wild. 2021. "Cartesian Frames." arXiv. https://doi.org/10.48550/arXiv.2109.10996.

Gemp, Ian, Thomas Anthony, Yoram Bachrach, Avishkar Bhoopchand, Kalesha Bullard, Jerome Connor, Vibhavari Dasagi, et al. 2022. "Developing, Evaluating and Scaling Learning Agents in Multi-Agent Environments." arXiv. https://doi.org/10.48550/arXiv.2209.10958.

Gleave, Adam. 2023. "AI Safety in a World of Vulnerable Machine Learning Systems." FAR AI. March 5, 2023. https://far.ai/post/2023-03-safety-vulnerable-world/.

Google DeepMind. 2024. "AI achieves silver-medal standard solving International Mathematical Olympiad problems." July 31, 2024. https://deepmind.google/discover/blog/ai-solves-imo-problems-at-silver-medal-level/.

Greenblatt, Ryan, Buck Shlegeris, Kshitij Sachan, and Fabien Roger. 2024. "AI Control: Improving Safety Despite Intentional Subversion." arXiv. https://doi.org/10.48550/arXiv.2312.06942.

Gruetzemacher, Ross, Kyle Kilian, Iyngkarran Kumar, David Manheim, Mark Bailey, Pierre Harter, and José Hernández-Orallo. 2024. "Envisioning a Thriving Ecosystem for Testing & Evaluating Advanced AI." NIST AI EO Comment. https://www.nist.gov/system/files/documents/2024/02/15/ID019%20-%202024-02-02%2

0Wichita%20State%20University%20et.%20al%2C%20Comments%20on%20AI%20EO%20RFI.pdf.

Guest, Oliver, Michael Aird, and Seán Ó hÉigeartaigh. 2023. "Safeguarding the Safeguards: How Best to Promote Alignment in the Public Interest." Institute for AI Policy; Strategy. https://www.iaps.ai/research/safeguarding-the-safeguards.

Hadfield-Menell, Dylan, Anca Dragan, Pieter Abbeel, and Stuart Russell. 2024. "Cooperative Inverse Reinforcement Learning." arXiv. https://doi.org/10.48550/arXiv.1606.03137.

Hadshar, Rose. 2023. "A Review of the Evidence for Existential Risk from AI via Misaligned Power-Seeking." arXiv. https://doi.org/10.48550/arXiv.2310.18244.

Hammond, Lewis. 2023. "Multi-Agent Risks from Advanced AI." In Multi-Agent Security Workshop: Security as Key to AI Safety. https://neurips.cc/virtual/2023/82192.

Hassija, Vikas, Vinay Chamola, Atmesh Mahapatra, Abhinandan Singal, Divyansh Goel, Kaizhu Huang, Simone Scardapane, et al. 2024. "Interpreting Black-Box Models: A Review on Explainable Artificial Intelligence." Cognitive computation, 45–74. https://doi.org/10.1007/s12559-023-10179-8.

Hayes, Jamie, Ilia Shumailov, Eleni Triantafillou, Amr Khalifa, and Nicolas Papernot. 2024. "Inexact Unlearning Needs More Careful Evaluations to Avoid a False Sense of Privacy." arXiv. https://doi.org/10.48550/arXiv.2403.01218.

Heikkilä, Melissa. 2024. "The AI Act is done. Here's what will (and won't) change." MIT Technology Review. March 19, 2024. https://www.technologyreview.com/2024/03/19/1089919/the-ai-act-is-done-heres-what-will-and-wont-change/.

Hendrycks, Dan, Nicholas Carlini, John Schulman, and Jacob Steinhardt. 2022. "Unsolved Problems in ML Safety." arXiv. https://doi.org/10.48550/arXiv.2109.13916.

Hendrycks, Dan, and Mantas Mazeika. 2022. "X-Risk Analysis for AI Research." arXiv. https://doi.org/10.48550/arXiv.2206.05862.

Hubinger, Evan, Carson Denison, Jesse Mu, Mike Lambert, Meg Tong, Monte MacDiarmid, Tamera Lanham, et al. 2024. "Sleeper Agents: Training Deceptive LLMs That Persist Through Safety Training." arXiv. https://doi.org/10.48550/arXiv.2401.05566.

Information Commissioner's Office, and The Alan Turing Institute. 2022. "Explaining Decisions Made with AI." ICO. https://ico.org.uk/for-organisations/uk-gdpr-guidance-and-resources/artificial-intelligence/explaining-decisions-made-with-artificial-intelligence.

International Dialogues on AI Safety. 2024. "Beijing Statement." March 11, 2024. https://idais.ai/.

Irving, Geoffrey, Paul Christiano, and Dario Amodei. 2018. "AI Safety via Debate." arXiv. https://doi.org/10.48550/arXiv.1805.00899.

Ji, Jiaming, Tianyi Qiu, Boyuan Chen, Borong Zhang, Hantao Lou, Kaile Wang, Yawen Duan, et al. 2024. "AI Alignment: A Comprehensive Survey." arXiv. https://doi.org/10.48550/arXiv.2310.19852.

Juliussen, Bjørn Aslak, Jon Petter Rui, and Dag Johansen. 2023. "Algorithms That Forget: Machine Unlearning and the Right to Erasure." Computer Law & Security Review 51 (November):105885. https://doi.org/10.1016/j.clsr.2023.105885.

Kästner, Lena, and Barnaby Crook. 2023. "Explaining AI Through Mechanistic Interpretability." PhilSci. https://philsci-archive.pitt.edu/22747/.

Kenton, Zachary, Ramana Kumar, Sebastian Farquhar, Jonathan Richens, Matt MacDermott, and Tom Everitt. 2022. "Discovering Agents." arXiv. https://arxiv.org/abs/2208.08345v2.

Khlaaf, Heidy. 2023. "Toward Comprehensive Risk Assessments and Assurance of AI-Based Systems." Trail of Bits. https://www.trailofbits.com/documents/Toward_comprehensive_risk_assessments.pdf.

Kinniment, Megan, Lucas Jun Koba Sato, Haoxing Du, Brian Goodrich, Max Hasin, Lawrence Chan, Luke Harold Miles, et al. 2024. "Evaluating Language-Model Agents on Realistic Autonomous Tasks." arXiv. https://doi.org/10.48550/arXiv.2312.11671.

Knight, Will. 2024. "OpenAI's Long-Term AI Risk Team Has Disbanded." Wired. May 17, 2024. https://www.wired.com/story/openai-superalignment-team-disbanded/.

Krakovna, Victoria, and Janos Kramar. 2023. "Power-Seeking Can Be Probable and Predictive for Trained Agents." arXiv. https://doi.org/10.48550/arXiv.2304.06528.

Lanham, Tamera, Anna Chen, Ansh Radhakrishnan, Benoit Steiner, Carson Denison, Danny Hernandez, Dustin Li, et al. 2023. "Measuring Faithfulness in Chain-of-Thought Reasoning." arXiv. https://doi.org/10.48550/arXiv.2307.13702.

Leibo, Joel Z., Edgar Duéñez-Guzmán, Alexander Sasha Vezhnevets, John P. Agapiou, Peter Sunehag, Raphael Koster, Jayd Matyas, Charles Beattie, Igor Mordatch, and Thore Graepel. 2021. "Scalable Evaluation of Multi-Agent Reinforcement Learning with Melting Pot." arXiv. https://doi.org/10.48550/arXiv.2107.06857.

Leike, Jan, and Ilya Sutskever. 2023. "Introducing Superalignment." OpenAI. July 5, 2023. https://openai.com/blog/introducing-superalignment.

Li, Nathaniel, Alexander Pan, Anjali Gopal, Summer Yue, Daniel Berrios, Alice Gatti, Justin D. Li, et al. 2024 "The WMDP Benchmark: Measuring and Reducing Malicious Use With Unlearning." arXiv. https://doi.org/10.48550/arXiv.2403.03218.

Lin, Stephanie, Jacob Hilton, and Owain Evans. 2022. "TruthfulQA: Measuring How Models Mimic Human Falsehoods." arXiv. https://doi.org/10.48550/arXiv.2109.07958.

Lu, Chris, Cong Lu, Robert Tjarko Lange, Jakob Foerster, Jeff Clune, and David Ha. 2024. "The AI Scientist: Towards Fully Automated Open-Ended Scientific Discovery." arXiv. https://doi.org/10.48550/arXiv.2408.06292.

MacCarthy, Mark. 2023. "The US and Its Allies Should Engage with China on AI Law and Policy." Brookings. October 19, 2023. https://www.brookings.edu/articles/the-us-and-its-allies-should-engage-with-china-on-ai-law-and-policy/.

Mandelbaum, Eric. 2022. "Everything and More: The Prospects of Whole Brain Emulation." Journal of Philosophy 119 (8): 444–59. https://doi.org/10.5840/jphil2022119830.

Matthews, Dylan. 2023. "The $1 Billion Gamble to Ensure AI Doesn't Destroy Humanity." Vox. September 25, 2023. https://www.vox.com/future-perfect/23794855/anthropic-ai-openai-claude-2.

Metz, Cade. 2016. "Inside OpenAI, Elon Musk's Wild Plan to Set Artificial Intelligence Free." Wired, April. https://www.wired.com/2016/04/openai-elon-musk-sam-altman-plan-to-set-artificial-intelligence-free/.

Metz, Rachel. 2024. "Ex-OpenAI Safety Leader Leike to Join Rival Anthropic." Bloomberg. May 28, 2024. https://www.bloomberg.com/news/articles/2024-05-28/ex-openai-safety-leader-leike-to-join-rival-anthropic.

Mukobi, Gabriel. 2024. "Reasons to Doubt the Impact of AI Risk Evaluations." arXiv. https://doi.org/10.48550/arXiv.2408.02565.

Murgia, Madhumita. 2023. "Google's DeepMind-Brain Merger: Tech Giant Regroups for AI Battle." Financial Times, April. https://www.ft.com/content/f4f73815-6fc2-4016-bd97-4bace459e95e.

Nannini, Luca, Agathe Balayn, and Adam Leon Smith. 2023. "Explainability in AI Policies: A Critical Review of Communications, Reports, Regulations, and Standards in the EU, US, and UK." arXiv. https://doi.org/10.48550/arXiv.2304.11218.

NIST. 2024. "Strategic Vision." NIST, May. https://www.nist.gov/aisi/strategic-vision.

Ngo, Richard, Lawrence Chan, and Sören Mindermann. 2024. "The Alignment Problem from a Deep Learning Perspective." arXiv. https://doi.org/10.48550/arXiv.2209.00626.

O'Brien, Joe, Shaun Ee, and Zoe Williams. 2023. "Deployment corrections: An incident response framework for frontier AI models." Institute for AI Policy and Strategy. https://www.iaps.ai/research/deployment-corrections.

Olah, Chris. 2023. "Interpretability Dreams." Transformer Circuits. May 24, 2023. https://transformer-circuits.pub/2023/interpretability-dreams/index.html.

OpenAI. 2023. "Preparedness." December 18, 2023. https://openai.com/safety/preparedness.

Open Philanthropy. 2022. "Request for Proposals for Projects in AI Alignment That Work with Deep Learning Systems. Open Philanthropy." January 2022. https://www.openphilanthropy.org/request-for-proposals-for-projects-in-ai-alignment-that-work-with-deep-learning-systems/.

Ouyang, Long, Jeff Wu, Xu Jiang, Diogo Almeida, Carroll L. Wainwright, Pamela Mishkin, Chong Zhang, et al. 2022. "Training Language Models to Follow Instructions with Human Feedback." arXiv. https://doi.org/10.48550/arXiv.2203.02155.

Pacchiardi, Lorenzo, Alex J. Chan, Sören Mindermann, Ilan Moscovitz, Alexa Y. Pan, Yarin Gal, Owain Evans, and Jan Brauner. 2023. "How to Catch an AI Liar: Lie Detection in Black-Box LLMs by Asking Unrelated Questions." arXiv. https://doi.org/10.48550/arXiv.2310.01405.

Pan, Alexander, Jun Shern Chan, Andy Zou, Nathaniel Li, Steven Basart, Thomas Woodside, Jonathan Ng, Hanlin Zhang, Scott Emmons, and Dan Hendrycks. 2023. "Do the Rewards Justify the Means? Measuring Trade-Offs Between Rewards and Ethical Behavior in the MACHIAVELLI Benchmark." arXiv. https://doi.org/10.48550/arXiv.2304.03279.

Park, Peter S., Simon Goldstein, Aidan O'Gara, Michael Chen, Dan Hendrycks. 2024. "AI deception: A survey of examples, risks, and potential solutions." Patterns. https://doi.org/10.1016/j.patter.2024.100988.

Pauketat, Janet VT, Justin Bullock, and Jacy Reese Anthis. 2023. "Public Opinion on AI Safety: AIMS 2023 Supplement." OSF. https://doi.org/10.31234/osf.io/jv9rz.

Perez, Ethan, Sam Ringer, Kamilė Lukošiūtė, Karina Nguyen, Edwin Chen, Scott Heiner, Craig Pettit, et al. 2022. "Discovering Language Model Behaviors with Model-Written Evaluations." arXiv. https://doi.org/10.48550/arXiv.2212.09251.

Perrigo, Billy. 2023a. "DeepMind CEO Demis Hassabis Urges Caution on AI." TIME, January. https://time.com/6246119/demis-hassabis-deepmind-interview/.

———. 2023b. "Bing's AI Is Threatening Users. That's No Laughing Matter." TIME, February. https://time.com/6256529/bing-openai-chatgpt-danger-alignment/.

Piper, Kelsey. 2019. "StarCraft is a deep, complicated war strategy game. Google's AlphaStar AI crushed it." Vox. January 25, 2019.

https://www.vox.com/future-perfect/2019/1/24/18196177/ai-artificial-intelligence-google-deepmind-starcraft-game.

Poli, Michael, Stefano Massaroli, Eric Nguyen, Daniel Y. Fu, Tri Dao, Stephen Baccus, Yoshua Bengio, Stefano Ermon, and Christopher Ré. 2023. "Hyena Hierarchy: Towards Larger Convolutional Language Models." arXiv. https://doi.org/10.48550/arXiv.2302.10866.

Pouget, Hadrien. 2024. "What will the role of standards be in AI governance?" Ada Lovelace Institute. April 5, 2023. https://www.adalovelaceinstitute.org/blog/role-of-standards-in-ai-governance/.

Radhakrishnan, Ansh, Karina Nguyen, Anna Chen, Carol Chen, Carson Denison, Danny Hernandez, Esin Durmus, et al. 2023. "Question Decomposition Improves the Faithfulness of Model-Generated Reasoning." arXiv. https://doi.org/10.48550/arXiv.2307.11768.

Räuker, Tilman, Anson Ho, Stephen Casper, and Dylan Hadfield-Menell. 2023. "Toward Transparent AI: A Survey on Interpreting the Inner Structures of Deep Neural Networks." In, 464–83. IEEE Computer Society. https://doi.org/10.1109/SaTML54575.2023.00039.

Reuel, Anka, Ben Bucknall, Stephen Casper, Tim Fist, Lisa Soder, Onni Aarne, Lewis Hammond, et al. 2024. "Open Problems in Technical AI Governance." arXiv. https://doi.org/10.48550/arXiv.2407.14981.

Rudner, Tim G. J., and Helen Toner. 2024. "Key Concepts in AI Safety: Reliable Uncertainty Quantification in Machine Learning." Center for Security and Emerging Technology. https://cset.georgetown.edu/publication/key-concepts-in-ai-safety-reliable-uncertainty-quantification-in-machine-learning/.

Russell, Stuart J. 2019. Human Compatible: Artificial Intelligence and the Problem of Control. New York: Viking.

Scheurer, Jérémy, Mikita Balesni, and Marius Hobbhahn. 2023. "Technical Report: Large Language Models Can Strategically Deceive Their Users When Put Under Pressure." arXiv. https://doi.org/10.48550/arXiv.2311.07590.

Schiermeier, Quirin. 2015. "The science behind the Volkswagen emissions scandal." Nature. September 24, 2015. https://www.nature.com/articles/nature.2015.18426.

Schuett, Jonas, Noemi Dreksler, Markus Anderljung, David McCaffary, Lennart Heim, Emma Bluemke, and Ben Garfinkel. 2023. "Towards Best Practices in AGI Safety and Governance: A Survey of Expert Opinion." arXiv. https://doi.org/10.48550/arXiv.2305.07153.

Schwartz, Victor. 1998. "The 'Restatement (third) of Torts: Products Liability': A Guide to Its Highlights." Tort & Insurance Law Journal. 31 (1): 85-100. http://www.jstor.org/stable/25763264.

Sharkey, Lee, Clíodhna Ní Ghuidhir, Dan Braun, Jérémy Scheurer, Mikita Balesni, Lucius Bushnaq, Charlotte Stix, and Marius Hobbhahn. 2023. "A Causal Framework for AI Regulation and Auditing." Apollo Research. https://www.apolloresearch.ai/research/a-causal-framework-for-ai-regulation-and-auditing.

Shevlane, Toby, Sebastian Farquhar, Ben Garfinkel, Mary Phuong, Jess Whittlestone, Jade Leung, Daniel Kokotajlo, et al. 2023. "Model Evaluation for Extreme Risks." arXiv. https://doi.org/10.48550/arXiv.2305.15324.

Soares, Nate, and Benya Fallenstein. 2017. "Agent Foundations for Aligning Machine Intelligence with Human Interests: A Technical Research Agenda." In The Technological Singularity: Managing the Journey, edited by Victor Callaghan, James Miller, Roman Yampolskiy, and Stuart Armstrong, 103–25. The Frontiers Collection. Berlin, Heidelberg: Springer. https://doi.org/10.1007/978-3-662-54033-6_5.

Stewart, Harlan. 2024. "June 2024 Newsletter." Machine Intelligence Research Institute. June 14, 2024. https://intelligence.org/2024/06/14/june-2024-newsletter/.

Templeton, Adly, Tom Conerly, Jonathan Marcus, Jack Lindsey, Trenton Bricken, Brian Chen, Adam Pearce, et al. 2024. "Scaling Monosemanticity: Extracting Interpretable Features from Claude 3 Sonnet." Transformer Circuits Thread. https://transformer-circuits.pub/2024/scaling-monosemanticity/index.html.

The Economist. 2014. "The dozy watchdogs." December 11, 2014. https://www.economist.com/briefing/2014/12/11/the-dozy-watchdogs.

The White House. 2023. "Executive Order on the Safe, Secure, and Trustworthy Development and Use of Artificial Intelligence." October 30, 2023. https://www.whitehouse.gov/briefing-room/presidential-actions/2023/10/30/executive-order-on-the-safe-secure-and-trustworthy-development-and-use-of-artificial-intelligence/.

Toner, Helen, and Ashwin Acharya. 2022. "Exploring Clusters of Research in Three Areas of AI Safety." Center for Security and Emerging Technology. https://cset.georgetown.edu/publication/exploring-clusters-of-research-in-three-areas-of-ai-safety/.

UK AI Safety Institute. 2024a. "Advanced AI Evaluations at AISI: May Update | AISI Work." May 20, 2024. https://www.aisi.gov.uk/work/advanced-ai-evaluations-may-update.

———. 2024b. "Systemic AI Safety Fast Grants." 2024. https://www.aisi.gov.uk/grants.

Vinitsky, Eugene, Raphael Köster, John P Agapiou, Edgar A Duéñez-Guzmán, Alexander S Vezhnevets, and Joel Z Leibo. 2023. "A Learning Agent That Acquires Social Norms from Public Sanctions in Decentralized Multi-Agent Settings." Collective Intelligence 2 (2): 26339137231162025. https://doi.org/10.1177/26339137231162025.

Yudkowsky, Eliezer. 2018. "The Rocket Alignment Problem. Machine Intelligence Research Institute." October 3, 2018. https://intelligence.org/2018/10/03/rocket-alignment/.

Zhang, Yue, Yafu Li, Leyang Cui, Deng Cai, Lemao Liu, Tingchen Fu, Xinting Huang, et al. 2023. "Siren's Song in the AI Ocean: A Survey on Hallucination in Large Language Models." arXiv. https://doi.org/10.48550/arXiv.2309.01219.

Ziegler, Daniel M., Seraphina Nix, Lawrence Chan, Tim Bauman, Peter Schmidt-Nielsen, Tao Lin, Adam Scherlis, et al. 2022. "Adversarial Training for High-Stakes Reliability." arXiv. https://doi.org/10.48550/arXiv.2205.01663.

Zimmermann, Roland S., Thomas Klein, and Wieland Brendel. 2024. "Scale Alone Does Not Improve Mechanistic Interpretability in Vision Models." Advances in Neural Information Processing Systems 36 (February). https://proceedings.neurips.cc/paper_files/paper/2023/hash/b4aadf04d6fde46346db455402860708-Abstract-Conference.html.

Zou, Andy, Long Phan, Sarah Chen, James Campbell, Phillip Guo, Richard Ren, Alexander Pan, et al. 2023a. "Representation Engineering: A Top-Down Approach to AI Transparency." arXiv. https://doi.org/10.48550/arXiv.2310.01405.

Zou, Andy, Zifan Wang, Nicholas Carlini, Milad Nasr, J. Zico Kolter, and Matt Fredrikson. 2023b. "Universal and Transferable Adversarial Attacks on Aligned Language Models." arXiv. https://doi.org/10.48550/arXiv.2307.15043.